\documentclass[journal=apchd5,manuscript=article]{achemso}

\usepackage[version=3]{mhchem}
\usepackage{graphicx}
\usepackage{amsmath,amssymb}
\usepackage{upgreek}
\usepackage{units}
\usepackage{booktabs}
\usepackage{xcolor}

\global\long\def\ket#1{\left|#1\right\rangle }%
\global\long\def\bra#1{\left\langle #1\right|}%

\title{Aspect ratio controls hot-carrier generation in gold nanobricks}
\author{Sim\~ao M. Jo\~ao}
\affiliation{Department of Materials, Imperial College London, South Kensington Campus, London SW7 2AZ, United Kingdom}

\author{Ottavio Bassano}
\affiliation{Department of Materials, Imperial College London, South Kensington Campus, London SW7 2AZ, United Kingdom}

\author{Johannes Lischner}
\affiliation{Department of Materials, Imperial College London, South Kensington Campus, London SW7 2AZ, United Kingdom}
\alsoaffiliation{The Thomas Young Centre for Theory and Simulation of Materials, London E1 4NS, United Kingdom}
\email{j.lischner@imperial.ac.uk}

\keywords{Nanoparticles, hot carriers, plasmonics, photocatalysis}
\begin{document}
\maketitle


\begin{abstract}
Energetic or "hot" electrons and holes generated from the decay of localized surface plasmons in metallic nanoparticles have great potential for applications in photocatalysis, photovoltaics, and sensing. Here, we study the generation of hot carriers in brick-shaped gold nanoparticles using a recently developed modelling approach that combines a solution to Maxwell's equation with large-scale tight-binding simulations to evaluate Fermi's Golden Rule. We find that hot-carrier generation depends sensitively on the aspect ratio of the nanobricks with flatter bricks producing a large number of energetic electrons irrespective of the light polarization. In contrast, the hot-carrier generation rates of elongated nanobricks exhibits a strong dependence on the light polarization. The insights resulting from our calculations can be harnessed to design nanobricks that produce hot carriers with properties tailored to specific device applications.
\end{abstract}

\section{Introduction}
As the global economy transitions towards a more sustainable future, research efforts are increasingly focused on developing innovative methods to harness solar energy. One promising avenue involves hot carriers, which are highly energetic electrons and holes generated from the absorption of sunlight. Despite their short lifetimes, hot carriers can be harnessed for a range of applications, including photocatalysis, photovoltaics, sensing, and optoelectronics\cite{konig2010hot, brongersma_plasmon-induced_2015, mukherjee_hot_2013, cortes_challenges_2020}. Advancing our understanding of how to effectively generate and utilize hot carriers is a crucial step towards designing highly efficient solar energy conversion devices.

Plasmonic nanoparticles exhibit a very strong and highly tunable interaction with light caused by localized surface plasmons (LSPs)~\cite{maier_plasmonics_2007}. When the LSP decays via the Landau damping mechanism, hot carriers are generated~\cite{maier_plasmonics_2007,khurgin2019hot,dubi2019hot,sundararaman2014theoretical}. The properties of both the LSP and the hot carriers are highly tunable and depend sensitively on the material composition, shape, size and environment of the nanoparticle. For example, Manjavacas and coworkers~\cite{manjavacas_plasmon-induced_2014} used a spherical well model to study properties of hot carriers in spherical Ag nanoparticles and found that the efficiency of hot-carrier generation depends sensitively on the nanoparticle diameter. More recently, Jin et al.~\cite{jin_plasmon-induced_2022} used an atomistic modelling technique to study spherical nanoparticles of Ag, Au and Cu and investigated the dependence of hot-carrier generation from intraband and interband transitions on the nanoparticle size and dielectric environment. 

In contrast to spherical nanoparticles, hot-generation generation in nanoparticles with non-spherical shapes is less well understood. For example, Zhang and Govorov~\cite{zhang_optical_2014} used a particle-in-a-box model to study hot-carrier properties in Au cubes and also in thin Au slabs. Santiago and coworkers analyzed hot-carrier generation in Au nanorods and also in different Au nanostars and demonstrated a strong dependence of the hot-carrier generation efficiency on the nanoparticle shape~\cite{santiago2020efficiency}. Very recently, Kang et al. analyzed hot-carrier generation in Au nanocubes, Au octahedra and also Au rhombic dodecahedra and found that nanocubes and octahedra generate significantly more hot carriers than the dodecahedra~\cite{kang_effect_2024}.

In this paper, we present a systematic study of hot-carrier generation in Au nanobricks with a square base and different heights. Such nanobricks feature sharp edges and corners which can act as hot spots for light absorption~\cite{langbein1976normal} and hot-carrier generation. To obtain hot-carrier generation rates, we use a recently developed atomistic technique to evaluate Fermi's golden rule~\cite{jin_plasmon-induced_2022,joao_atomistic_2023}. The numerical cost of this approach increases only linearly with the nanoparticle size allowing us to model nanobricks consisting of thousands of atoms. We find that hot-carrier properties depend sensitively on the aspect ratio of the nanobricks as well as on the light polarization: nanobricks with larger heights produce mostly hot holes when the electric field is perpendicular to the square base, while flatter nanobricks generate mostly hot electrons regardless of the electric field orientation. The insights from our study provide a mechanistic understanding of hot-carrier generation in Au nanobricks and can be used to design nanobricks for hot-carrier devices for specific applications in photocatalysis or sensing.

\section{Results and discussion}
We first analyse the electric field inside the nanobricks which is responsible for the excitation of hot electrons and holes. Then, results for the energetic distributions and excitation rate of hot carriers are presented.

\subsection{Field enhancement and optical absorption}
\begin{figure}[H]
    \centering
    \includegraphics[width=0.7\linewidth]{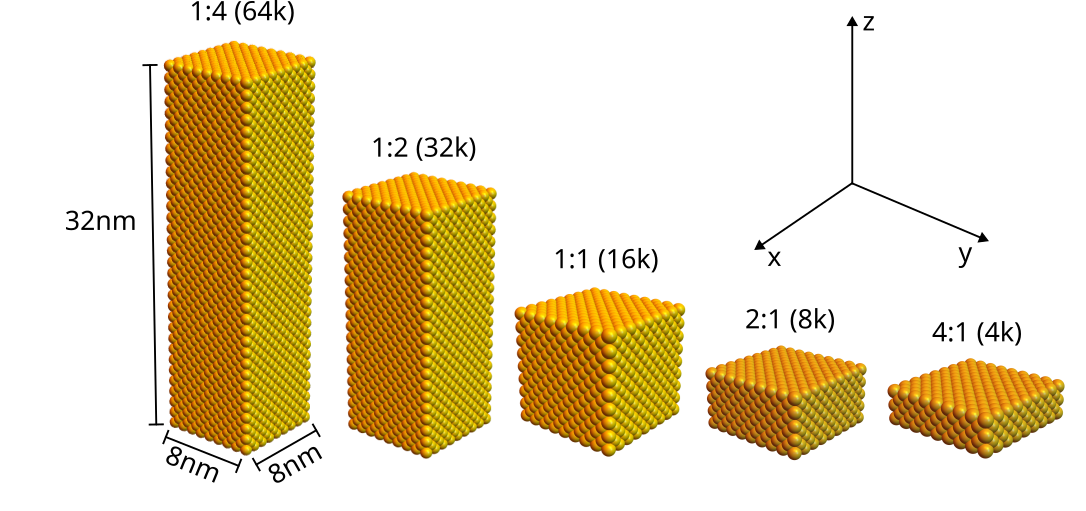}
    \caption{Geometries used in the simulations with corresponding aspect ratios, side lengths and number of atoms.}
    \label{fig:shapes}
\end{figure}

In this section, we analyse the electric field distribution in Au nanobricks of different aspect ratios illuminated by light polarized along the $x$ and $z$ directions (see Fig.~\ref{fig:shapes}). The Au nanobricks have a square base which lies in the $xy$-plane with a side length of $8$~nm and the aspect ratio is defined as the ratio of the base side length and the nanobrick height. We study aspect ratios ranging from 1:4 to 4:1, see Fig.~\ref{fig:shapes}, containing up to 64,000 Au atoms.

\begin{figure}[H]
    \centering
    \includegraphics[width=0.95\linewidth]{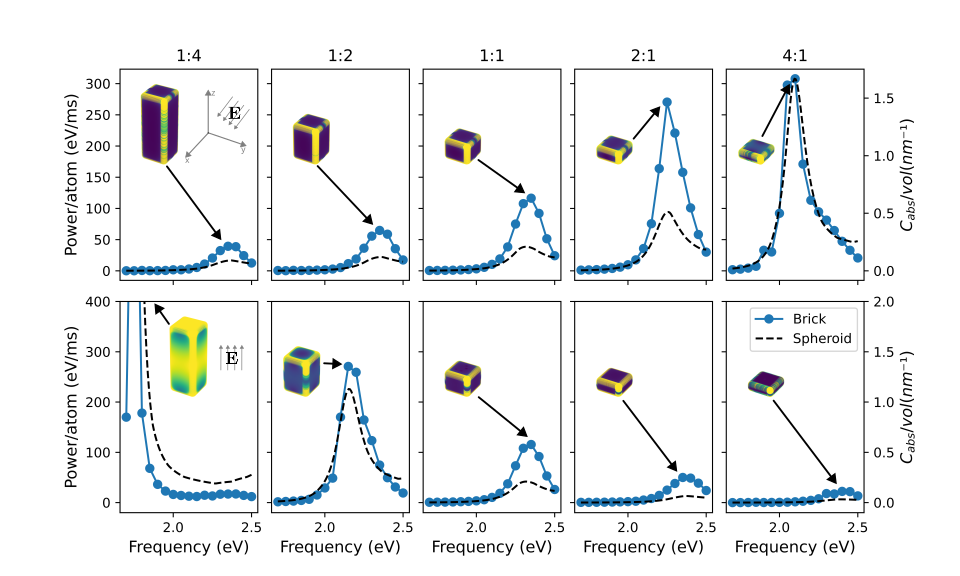}
    \caption{Power absorbed per atom as a function of illumination frequency for Au nanobricks of different aspect ratios (blue dots). In the top (bottom) panels, the electric field is polarized along the $x-$ ($z-$) axis. The electric field distribution at the LSPR frequency is shown as an inset in each panel. The black dashed curves show the absorption cross section of an Au spheroid with the same aspect ratio (with a red-shift of  $0.1$~eV).}
    \label{fig:LSPR}
\end{figure}

Figure~\ref{fig:LSPR} shows the power absorbed by nanobricks of aspect ratios ranging from 1:4 to 4:1 illuminated by light polarized along the $x$-direction (top panels) and the $z$-direction (bottom panels). The polarization-averaged results can be found in the Supplementary Material. The results are obtained by solving the Maxwell equations using the quasi-static approximation (see Methods for details). The illumination frequencies were chosen to range from 1.7 eV to 2.5 eV in order to cover the LSPR frequencies of all nanobricks under consideration. In each panel, the electric field distribution at the LSPR frequency is shown as an inset.

When the incident electric field is polarized along $x$, the resonant frequency shifts from 2.4 eV to 2.1 eV as the aspect ratio is changed from 1:4 to 4:1, i.e., as the nanobricks get flatter, the resonance red-shifts. This red-shift is accompanied
a significant increase in the absorbed power per atom (by approximately a factor of $7$). In contrast, when the incident light is polarized along $z$, the LSPR blue-shifts as the nanobricks get flatter, and the normalized power absorbed at the LSPR is $20$ times larger for an aspect ratio of 1:2 compared to 4:1. Most of these findings can be understood by approximating the nanobricks as prolate and oblate spheroids with the same aspect ratios (see Supplementary Information). The absorption cross section of such spheroids can be expressed in terms of a \textit{depolarization factor} $L_z$ whose magnitude is proportional to the aspect ratio: the smaller $L_z$, the easier it is to polarize the nanoparticle along the $z$-direction relative to $x$- or $y$-directions, causing a stronger field enhancement and a red-shift in the $z$-direction and a weaker field enhancement and a blue-shift for $x$- and $y$-directions.

Approximating nanobricks by spheroids of the same aspect ratio accurately reproduces the LSPR frequency and intensity, but it fails to predict the electric field distribution inside the nanobricks. Inside the spheroids, the electric field is uniform (although not necessarily parallel to the external field)~\cite{gans_uber_1912, bohren_absorption_2008}. In contrast, the electric field of the nanobricks is concentrated near the corners and edges. Even though we only show the distribution at the LSPR, we observe similar electric field distributions at other frequencies. Interestingly, the electric field is also strong in the middle of the long surface of the 4:1 nanobrick when the field is along $z$. This is likely a consequence of the presence of multiple plasmon modes: unlike nanospheres, nanocubes and nanobricks feature several plasmon resonances in the quasistatic approximation \cite{fuchs_theory_1975, fuchs_optical_1966, zhang_substrate-induced_2011} which give rise to field enhancements in different regions of the nanoparticle.

\subsection{Hot-carrier generation}
\begin{figure}[H]
    \centering
    \includegraphics[width=0.8\linewidth]{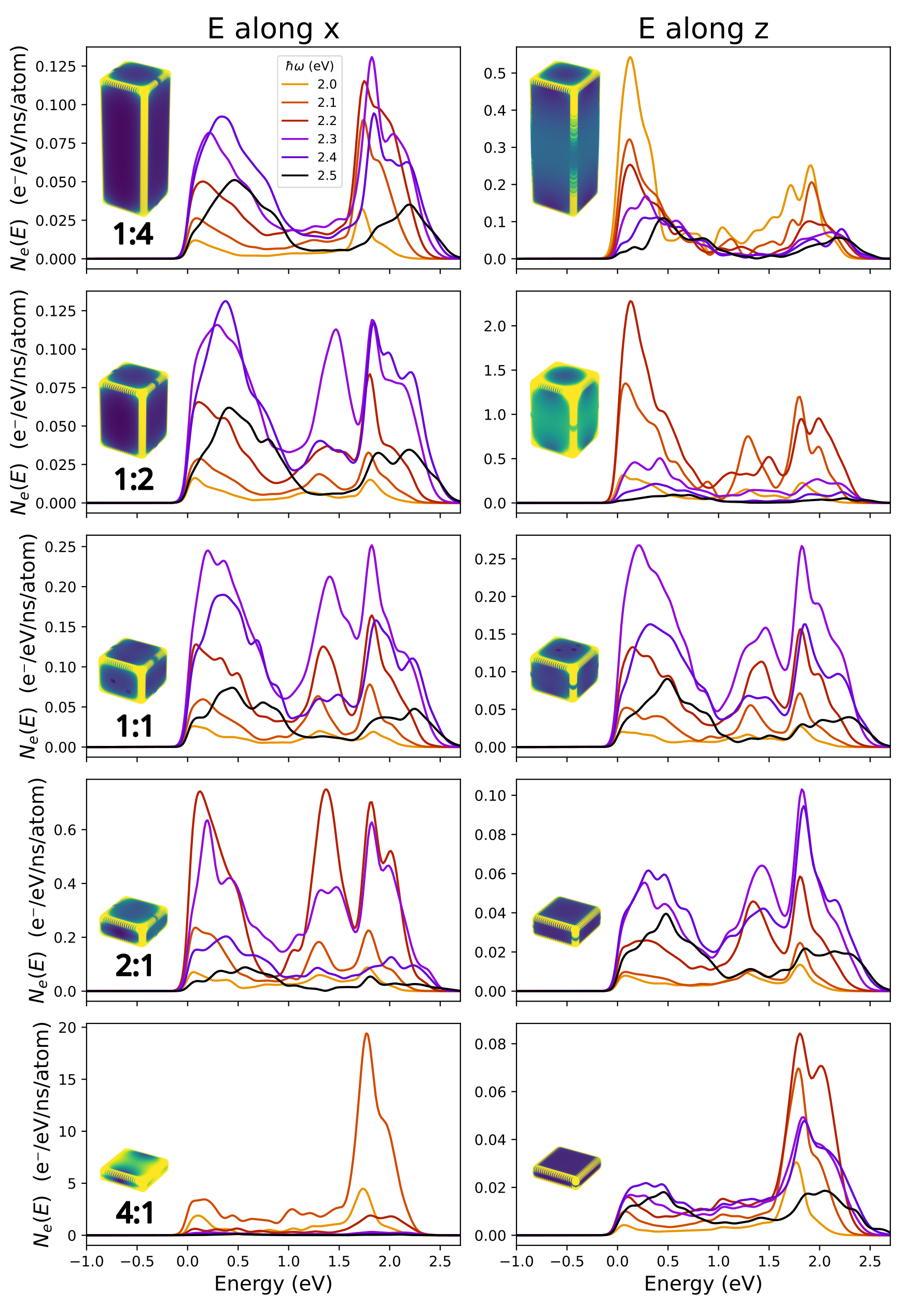}
    \caption{Electron generation rate of Au nanobricks with aspect ratios ranging from 1:4 to 4:1 and light frequencies ranging from 2.0 eV to 2.5 eV. The electric field on the left (right) column is polarized along the $x-$ ($z-$) axis. The insets correspond to the electric field distribution at the LSPR, and provide a visual representation of the overall intensity of the electric field in the NP.}
    \label{fig:grid2}
\end{figure}

Next, we study the properties of the electrons and holes which are excited by the time-dependent electric field. Fig.~\ref{fig:grid2} shows the hot-electron generation rates for Au nanobricks of different aspect ratios and different light frequencies and polarizations obtained from Fermi's Golden Rule (see Methods for details). Note that the hot-hole generation rates can be obtained from the hot-electron results simply by red-shifting the curves by $\hbar \omega$.

The hot-electron generation rates have contributions from two mechanisms: interband transitions from the occupied d-bands to unoccupied sp-band states generate energetic holes, but electrons with energies close to the Fermi level. In contrast, surface-enabled intraband transitions within the sp-band give rise to energetic electrons, but "cold" holes. For example, in the top left panel of Fig.~\ref{fig:grid2} it can be seen that the hot-electron generation rate of a 1:4 nanobrick (illuminated by light polarized along the $x$-direction) exhibits two peaks of approximately equal height with the low-energy peak corresponding to electrons generated from interband transitions and the high-energy peak to electrons produced by intraband transitions. 

As the nanobricks become flatter, a larger portion of electrons is generated via intraband transitions leading to a larger relative amount of hot electrons and a more pronounced intraband peak. This is a consequence of the larger surface-to-volume ratio of the flatter nanobricks. Comparing the two light polarizations, we find that, in general, there is a higher rate of carrier generation when the electric field is pointing along a long axis (e.g. the z-axis for the 1:4 aspect ratio and the x-axis for the 4:1 aspect ratio).

For device applications, such as in plasmonic catalysis, it is useful to introduce a simple metric to quantify the number of highly energetic carriers. For this, we define a threshold energy $E_t=1$~eV and define highly energetic electrons (holes) to have energies larger than $E_t$ (less than $-E_t$) relative to the Fermi level. Integrating the electron generation rates from Fig.~\ref{fig:grid2} over all the energies yields the total number of electrons $N^\text{tot}_e(\omega)$ being generated per unit time. Integrating only over energies larger than $E_t$ gives the number of highly energetic electrons generated per unit time $N_{he}(\omega)$. The total number of highly energetic holes $N_{hh}(\omega)$ generated per unit time is obtained by integrating the hole generation rate over all energies less than $-E_t$. Because of energy conversation, we can also obtain $N_{hh}(\omega)$ by integrating the electron generation rate from 0 to $\hbar\omega-E_t$, where $0$ is set as the Fermi level:

\begin{eqnarray*}
N_{\text{hh}}\left(\omega\right) & = & \int_{0}^{\hbar\omega-E_t}N_{e}\left(E,\omega\right)dE,\\
N_{\text{he}}\left(\omega\right) & = & \int_{E_t}^{\infty}N_{e}\left(E,\omega\right)dE.
\end{eqnarray*}

\begin{figure}[H]
    \centering
    \includegraphics[width=0.75\linewidth]{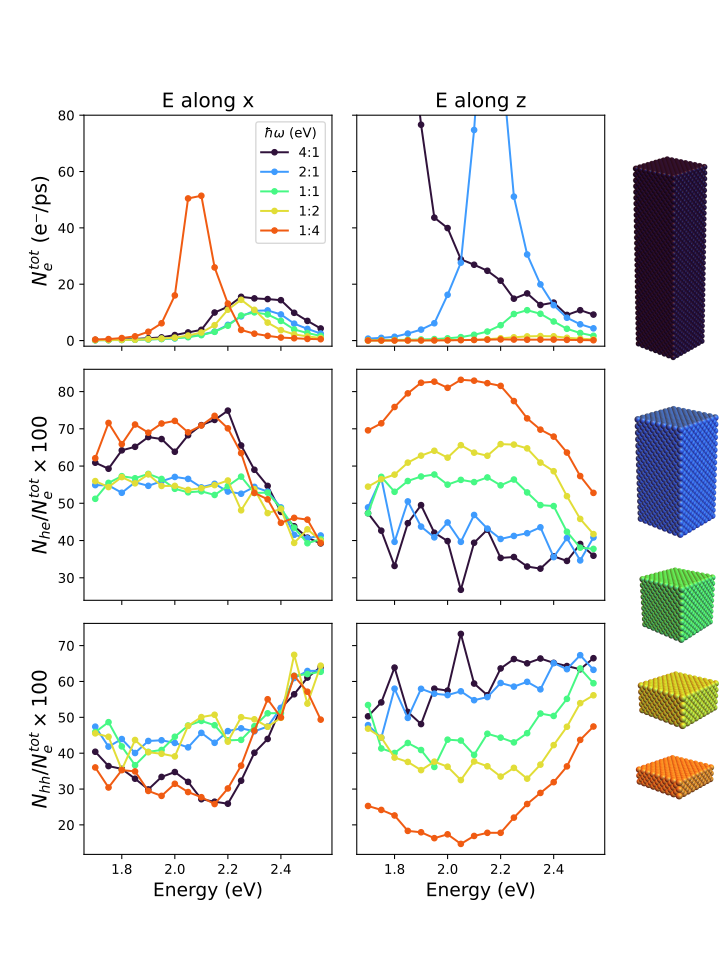}
    \caption{Total electron generation rate (top panels) and percentage of highly energetic electrons with energies larger than 1 eV relative to the Fermi level (middle panels) and highly energetic holes with energies less than -1 eV (bottom panels), as a function of the illumination frequency. The electric field is polarized along $x-$ ($z-$) direction in the left (right) panels. The curves are color-coded according to the nanobricks shown on the right.}
    \label{fig:hot_percentage}
\end{figure}

The top panels of Fig.~\ref{fig:hot_percentage} show the total number of electrons $N_e^\text{tot}$ generated upon illumination for each of the aspect ratios. The middle (bottom) panels present the percentage of highly energetic electrons (hot holes). The left (right) panels are obtained for an electric field polarized along the $x-$ ($z-$) direction.

As expected, the frequency dependence of $N_e^\text{tot}(\omega)$ is similar to that of the absorbed power, see Fig.~\ref{fig:LSPR}, exhibiting peaks at the LSPR frequencies of the different nanobricks. For light polarized along the $x$-direction, $N_e^\text{tot}$ is largest for flat nanobricks. In contrast, long nanobricks exhibit the largest peak in the total electron generation when the polarization is along the $z$-direction.

The middle and bottom panels show that the fraction of highly energetic holes $N_{hh}(\omega)$ (highly energetic electrons $N_{eh}(\omega)$) increases (decreases) for higher frequencies. This can be explained by the fact that more $d$-band states become accessible at higher photon energies which can accommodate energetic holes which increases the contribution from interband transitions. 

Comparing the results for the different light polarizations, we observe a different behaviour of the fraction of highly energetic holes and the fraction of highly energetic electrons: when the electric field is along the $x$-direction, the curves are very similar for all aspect ratios, but when the field is along the $z$-axis, the curves are offset from each other, even though they still follow a similar trend. This observation can be understood by considering the path of the electrons as they oscillate in response to the applied electric field. When the field is polarized along the $z$-direction, the electrons oscillate along trajectories which are parallel to the $z$-axis. If the height of the nanobrick is increased, the electrons propagate for longer before reaching a surface, effectively experiencing a more 'bulk-like' behaviour. In the bulk, intraband transitions are suppressed, so a higher fraction of hot holes is produced from interband transitions. In contrast, when the electric field is applied along the $x$-direction, the distance electrons can propagate before reaching a surface does not change as the aspect ratio changes and therefore the ratio of interband and intraband transitions remains approximately constant. Overall, flatter nanobricks produce more hot electrons than longer nanobricks because of the higher surface-to-volume ratio inducing a larger fraction of intraband transitions.

\section{Conclusion}
In this work, we analyzed the generation of hot carriers in Au nanobricks with a square base and variable height. We find that the optical absorption cross section of the nanobricks depends sensitively on their aspect ratio and is well reproduced by prolate and oblate spheroids of the same aspect ratios. This, however, is not the case for the electric field enhancement and the hot-carrier generation. All nanobricks produce a larger fraction of hot holes as the frequency of incident light increases since more $d$-band states below the Fermi level become energetically available: when the electric field is oriented along the $x$-direction (the square base of the nanobricks lies in the $xy$-plane) the fraction of hot holes increases from around 30\% to 70\% for all nanobricks as the frequency increases from 2.0 eV to 2.5 eV. In contrast, when the field is pointing along the $z$-direction, the fraction of hot electrons decreases from 80\% to 40\% when the height of the nanobricks increases as the electrons experience a more 'bulk-like' environment and intraband transitions are suppressed. These findings demonstrate the hot-carrier properties of Au nanobricks depend sensitively on their aspect ratio. This knowledge can be exploited for the design of hot-carrier devices for sensing and solar energy conversion.

\section{Methods}

\subsection{Atomic structure of gold nanobricks}

The atomic structure of the Au nanobricks was constructed by starting from an FCC lattice with a lattice constant $a=4.08$nm and retaining only those atoms which are inside the specific nanobrick shape.

\subsection{Electric field distribution}

The electric field distribution inside of the nanobricks is obtained by solving Maxwell's equation within the quasistatic approximation. This approximation produces accurate results when the wavelength of light is considerably larger than the size of the nanoparticle. In practice, we solve Laplace's equation for the total electric potential $\phi(\mathbf{r})$ (see \cite{jin_theory_2023} for an in-depth explanation):

\begin{equation}
\nabla\cdot\left(\varepsilon\left(\mathbf{r},\omega\right)\nabla\phi\left(\mathbf{r},\omega\right)\right)=0
\end{equation}
subject to the boundary condition $-\nabla\phi=\mathbf{E}_{0} $ at very large distances, where $\mathbf{E}_{0}$ is the external electric field with strength $1 \times 10^5$~V/m and $\epsilon\left(\mathbf{r},\omega\right)$ denotes the dielectric function~\cite{haynes_crc_2014}. Laplace's equation is solved numerically using the DC module in COMSOL Multiphysics.

The total power absorbed by the nanoparticle is given by \cite{jackson_classical_1998, bohren_absorption_2008}

$$ 
P_{\text{NP}}=\frac{\omega}{2}\text{Im}\left(\varepsilon\left(\omega\right)\right)\int_{V}\text{d}V\left|\mathbf{E}\left(\mathbf{r}, \omega\right)\right|^{2},
$$ 
where $\mathbf{E}\left(\mathbf{r},\omega\right)$ is the electric field, which is integrated over the volume $V$ of the nanoparticle.

\subsection{Electronic Hamiltonian and Hot-Carrier Generation}
Following Refs.\cite{joao_atomistic_2023,jin_plasmon-induced_2022,jin_theory_2023}, we construct the electronic Hamiltonian $\hat{H}_0$ of the nanoparticle using a two-center orthogonal Slater-Koster parametrization of the $5d$, $6s$ and $6p$ orbitals of Au \cite{papaconstantopoulos_handbook_2015}. The time-dependent electric potential described in the previous section induces electronic transitions from occupied orbitals in the nanobricks to empty orbitals. The time-dependent perturbation gives rise to an additional term in the Hamiltonian $\hat{H} = \hat{\Phi}\left({\omega}\right) + \hat{H}_0$ with

$$ \hat{\Phi}\left({\omega}\right)=\sum_{\mathbf{R}\alpha}\phi\left(\mathbf{R},\omega\right)\ket{\mathbf{R}\alpha}\bra{\mathbf{R}\alpha}. $$

The potential $\phi$ is being evaluated at the atomic positions $\mathbf{R}$, and $\ket{\mathbf{R}\alpha}$ denotes a tight-binding basis state at position $\mathbf{R}$ and orbital $\alpha$. The hot-electron generation rate is then computed using Fermi's Golden Rule to obtain the rate of optical transitions $\Gamma_{if}$ from initial occupied state $i$ to final unoccupied state $f$ 

\begin{equation}
    \Gamma_{if}(\omega)=\frac{2\pi}{\hbar} \Big|{\langle{\psi_f}| \hat{\Phi} (\omega)|\psi_i\rangle}\Big|^2 \delta(E_f-E_i-\hbar\omega;\sigma)f(E_i)(1-f(E_f)),
    \label{gamma_eqn}
\end{equation}

where $| \psi_{i} \rangle$ ($| \psi_{f} \rangle$) is the electronic wavefunction of the initial (final) state and $f\left(E\right)$ denotes the Fermi distribution at room temperature. Finally, the generation rate per unit volume of electrons with energy $E$ is obtained by summing up all the transitions that have final energy $E=E_f$

\begin{equation}
    N_e(E,\omega) = \frac{2}{V} \sum_{if} \Gamma_{if}(\omega) \delta(E-E_f;\sigma),
    \label{HCG_eqn}
\end{equation}

where $V$ is the nanoparticle volume and the factor of 2 accounts for spin degeneracy. We define $\delta(x;\sigma)=\frac{1}{\sqrt{2\pi\sigma^2}}\exp{(-\frac{x^2}{2\sigma^2})}$ which becomes a delta function in the limit of $\sigma\rightarrow0^+$ and $\sigma=0.06$~eV is a broadening parameter. The hole generation rate $N_h\left(E, \omega \right)$ is obtained by shifting the electron distribution $N_e\left(E, \omega \right)$ by $\hbar \omega$.

In practice, the previous expression is calculated using a basis-independent form of Fermi's Golden Rule 

\begin{equation}
N_{e}\left(E,\omega\right)=\frac{4\pi}{\hbar V}\int_{-\infty}^{\infty}d\varepsilon\delta\left(\varepsilon-E+\hbar\omega\right)f\left(\varepsilon\right)\left(1-f\left(E\right)\right)\text{Tr}\left[\delta\left(\varepsilon-\hat{H}_0\right)\hat{\Phi}\left({\omega}\right)\delta\left(E-\hat{H}_0\right)\hat{\Phi}^{\dagger}\left({\omega}\right)\right]
\end{equation}

which is then evaluated using the Kernel Polynomial Method \cite{weisse_chebyshev_2004,jin_plasmon-induced_2022}.

\subsection{Absorption cross-section of spheroids}
The power absorbed by the Au nanobricks can be understood from the analytical formula for the the absorption cross-section $C_{abs}$ of spheroids obtained from Gans' theory \cite{gans_uber_1912, bohren_absorption_2008}. The absorption cross-section 

\begin{equation}\label{eq:Cabs}
    C_{\text{abs}}=k\text{Im}\left(\sum_{i}\alpha_{i}\right)
\end{equation}

is expressed in terms of the polarization $\alpha_i$ induced by an electric field along the $i$ direction 

$$\alpha_{i}=4\pi a_{1}a_{2}a_{3}\frac{\varepsilon\left(\omega\right)-\varepsilon_{m}}{3\varepsilon_{m}+3L_{i}\left(\varepsilon\left(\omega\right)-\varepsilon_{m}\right)}.$$

Here, $k$ is the wavevector of the incident light, $\varepsilon_m$ is the dielectric constant of the medium surrounding the nanoparticle (set to $\varepsilon_m = \varepsilon_0$) and $a_1$, $a_2$ and $a_3$ are the lengths of the spheroid axes. The depolarization factor $L_i$ is given by 

$$ L_{i}=\frac{a_{1}a_{2}a_{3}}{2}\int_{0}^{\infty}\frac{dq}{\left(a_{i}^{2}+q\right)\sqrt{\left(q+a_{1}^{2}\right)\left(q+a_{2}^{2}\right)\left(q+a_{3}^{2}\right)}}. $$

Figure \ref{fig:LSPR} shows the volume-normalized absorption cross-section calculated with Eq.~\ref{eq:Cabs} for spheroids with $a_1 = a_2 = 8$~nm and $a_3$ varies from $0.25$ to $4.00$ in order to model the same aspect ratios as for the nanobricks.

\section{Data availability}
The data used for the findings in this study is available from the corresponding author upon reasonable request.

\section{Code availability}
The code used to generate the data in this study can be found on the public GitHub repository https://github.com/simaomenesesjoao/MaxTB.

\section{Author contributions}
S.M.J. and O.B. performed the simulations, J.L. supervised the work.

\section{Competing interests}
The authors declare no competing interests.

\begin{acknowledgement}
S.M.J. and J.L. acknowledge funding from the Royal Society through a Royal Society University Research Fellowship URF/R/191004 and also from the EPSRC programme grant EP/W017075/1. 
\end{acknowledgement}

\begin{suppinfo}
The supplementary information contains the absorption cross-sections for spheroids.
\end{suppinfo}

\bibliography{references}

\end{document}


\maketitle


\section{Polarization average}
Under the quasi-static approximation, the solution to Laplace's equation for a general external field $\mathbf{E}$ can be constructed from the solution to fields along the three Cartesian axes. Let $\Phi_x, \Phi_y, \Phi_z$ be the electric potential resulting from an electric field $\mathbf{E}_x, \mathbf{E}_y, \mathbf{E}_z$ along the $x,y,z$ directions respectively. Then, the electric potential induced by a generic field $\mathbf{E}= \alpha \mathbf{E}_x + \beta \mathbf{E}_y + \gamma \mathbf{E}_z$ is simply $\Phi = \alpha \Phi_x + \beta \Phi_y + \gamma \Phi_z$. Setting $\alpha, \beta, \gamma$ to be the direction cosines and $\mathbf{E}_x, \mathbf{E}_y, \mathbf{E}_z$ to be unit vectors, the potential is written as

$$ \Phi\left(\theta,\phi\right) = \sin\theta \cos\phi  \Phi_x + \sin\theta\sin\phi \Phi_y + \cos\theta  \Phi_z $$
and the polarization-averaged electron generation rate requires an average over the polar angle
$$N_{e}\left(E,\omega\right)\sim\int_{0}^{\pi}d\theta\int_{0}^{2\pi}d\phi \sin\theta\text{Tr}\left[\delta\left(\varepsilon-H_0\right)\Phi\left(\theta,\phi\right)\delta\left(E-H_0\right)\Phi^{\dagger}\left(\theta,\phi\right)\right] $$
The angular integrals can be solved analytically, showing that the average over polarization is identical to averaging over the three cartesian directions:
$$N_{e}\left(E,\omega\right)\sim\sum_{i=x,y,z}\text{Tr}\left[\Phi_{i}^{\dagger}\delta\left(\varepsilon-H_0\right)\Phi_{i}\delta\left(E-H_0\right)\right]$$

Furthermore, having defined $z$ as the long axis and $x,y$ as the short axes of the nanobricks, only the solutions along $x$ and $z$ are required since the nanobricks are symmetrical with respect to 90º rotations around the $z$ axis.

\begin{figure}[H]
    \centering
    \includegraphics[width=1\linewidth]{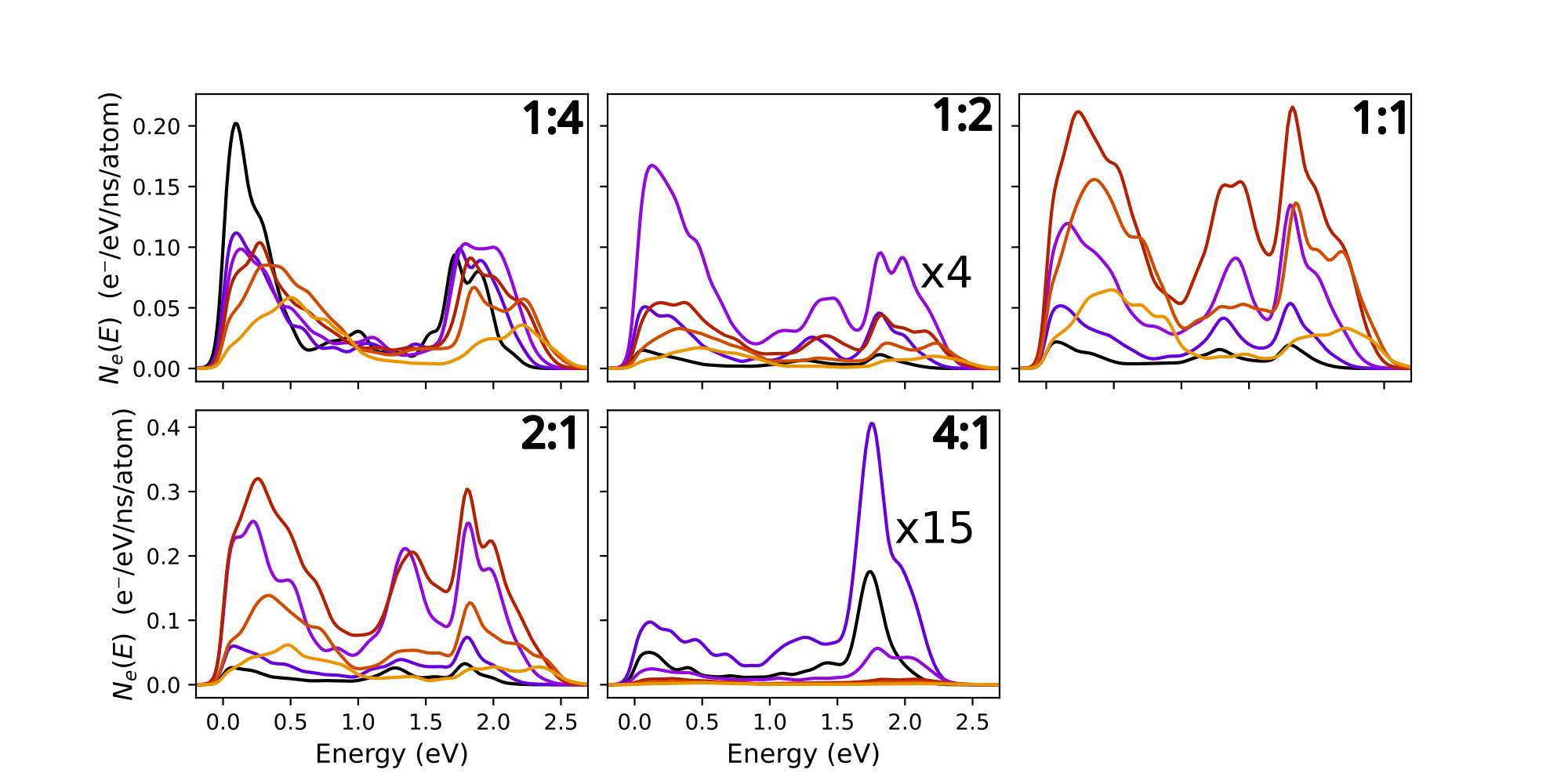}
    \caption{Electron generation rate for nanobricks with aspect ratios ranging from 1:4 to 4:1, averaged over all light polarizations. Each curve represents a different light frequency, from 2.0 eV to 2.5 eV.}
    \label{figsup:grid}
\end{figure}

Figure \ref{figsup:grid} represents the rate and energy at which electrons are being generated for different illumination frequencies and aspect ratios. This Figure is the polarization-averaged (or, equivalently, orientation-averaged) analogue of Fig. 2 in the main text and would be representative of the hot carrier generation rate of a distribution of randomly-oriented nanoparticles, such as in solution. Comparing to Fig. 2, the polarization averaged HCG for aspect ratios between 1:4 and 1:2 is dominated by interband transitions arising from the long nanoparticles with their axis aligned with the electric field. In contrast, the 2:1 and 4:1 aspect ratios are dominated by intraband transitions coming from flat nanoparticles whose short axis ($x$) is parallel to the electric field.

\begin{figure}[H]
    \centering
    \includegraphics[width=1\linewidth]{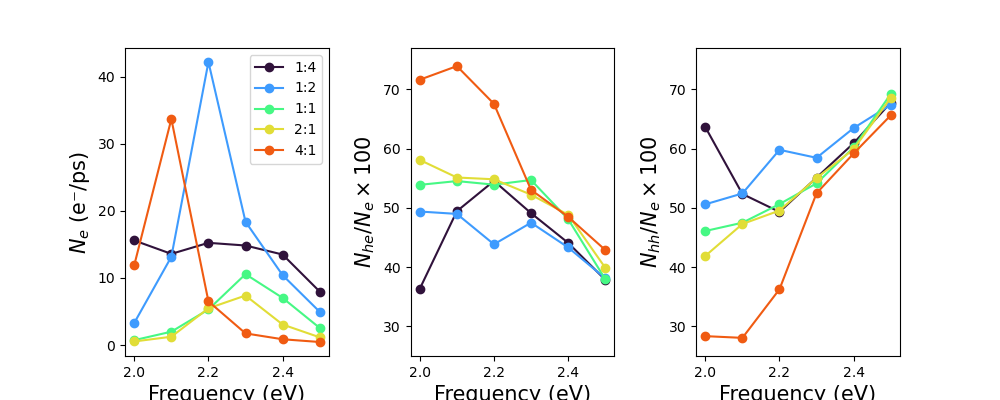}
    \caption{Total electron generation rate (left) and percentage of hot electrons (middle) and hot holes (right), as a function of illumination frequency averaged over all light polarizations. The curves are color-coded according to the main text.} 
    \label{figsup:hot_percentage}
\end{figure}

Figure \ref{figsup:hot_percentage} shows the total rate of electron generation $N_e$, as well as the ratio of hot electrons and holes to total electrons. $N_e$ attains the largest values for frequencies of $2.1$ and $2.2$ eV for the aspect ratios 4:1 and 1:2 respectively. As a general trend, the fraction of hot holes (electrons) tends to decrease (increase) as the frequency increases. The 4:1 AR shows the highest overall sensitivity in the proportion of hot electrons/holes to changes in frequency.